\documentclass[conference,a4paper]{IEEEtran}

\usepackage[cmex10]{amsmath}
\usepackage[cmex10]{amsmath}
\usepackage{array}
\usepackage{verbatim}
\usepackage{tikz}
\usepackage{epstopdf}
\usepackage{amsfonts}
\usepackage{amsmath}
\usepackage{amssymb}
\newtheorem{defi}{Definition}
\newtheorem{theorem}{Theorem}

\newtheorem{remark}{Remark}
\newtheorem{corollary}{Corollary}

\newcommand{\modn}{\hspace{-0.1in}\mod\Lambda}

\newcommand{\modvf}{\hspace{-0.1in}\mod\hspace{0.05in}\Lambda_{t,1}}

\newcommand{\moduf}{\hspace{-0.1in}\mod\hspace{0.05in}\Lambda_{u,1}}
\newcommand{\modus}{\hspace{-0.1in}\mod\hspace{0.05in}\Lambda_{u,2}}

\usepackage{cite}
\usepackage[center]{caption}
\usepackage{graphicx}
\usepackage{multicol}
\usetikzlibrary{shapes.misc}
\tikzset{cross/.style={cross out, draw=black, minimum size=2*(#1-\pgflinewidth), inner sep=0pt, outer sep=0pt},
cross/.default={1pt}}

\begin{document}
\title{Security in The Gaussian Interference Channel: Weak and Moderately Weak Interference Regimes }
\author{\IEEEauthorblockN{Parisa Babaheidarian*, Somayeh Salimi**, Panos Papadimitratos**}
\IEEEauthorblockA{
*Boston University,**KTH Royal Institute of Technology}}

\date{}
\maketitle
\begin{abstract}
We consider a secure communication scenario through the two-user Gaussian interference channel: each transmitter (user) has a confidential message to send reliably to its intended receiver while keeping it secret from the other receiver. Prior work investigated the performance of two different approaches for this scenario; i.i.d. Gaussian random codes and real alignment of structured codes. While the latter achieves the optimal sum secure degrees of freedom (s.d.o.f.), its extension to finite SNR regimes is challenging. In this paper, we propose a new achievability scheme for the weak and the moderately weak interference regimes, in which the reliability as well as the confidentiality of the transmitted messages are maintained at any finite SNR value. Our scheme uses lattice structure, structured jamming codewords, and lattice alignment in the encoding and the asymmetric compute-and-forward strategy in the decoding. We show that our lower bound on the sum secure rates scales linearly with $\log(SNR)$ and hence, it outperforms i.i.d. Gaussian random codes. Furthermore, we show that our achievable result is asymptotically optimal. Finally, we provide a discussion on an extension of our scheme to $K>2$ users.
\end{abstract}
\section{Introduction}
It has been shown that structured codes along with alignment techniques can improve achievability results over i.i.d. random codes in different secure communication scenarios. For instance, two schemes based on real alignment of structured signals were tested on a multi-user Gaussian interference channel with confidential messages \cite{xie2014secure}, \cite{xieJournal2}. In \cite{xie2014secure}, a two-user Gaussian interference channel with no helper was considered, in which each user had a message for its intended receiver to be kept confidential from the other receiver. It was shown that the optimal sum secure degrees of freedom can be achieved for this channel through cooperative jamming signals and real alignment \cite{xie2014secure}. The case with extra nodes serving as helpers was further investigated at infinite SNR in \cite{xie2014secure}. Also, the scheme proposed in \cite{xieJournal2} attained the optimal sum s.d.o.f. for the $K>2$-user Gaussian interference channels with confidential messages and no helper; it was shown that the optimal s.d.o.f. for this general case is equal to $\frac{K(K-1)}{2K-1}$ \cite{xieJournal2}.\\
\indent Although the aforementioned schemes showed promising performance in the infinite SNR regime, their extension to the finite SNR regimes is challenging due to the difficulty in bounding their decoding error probability at a finite SNR value. In this work, we consider the two-user Gaussian interference channel with no helper in which each transmitter wishes to send a message to its intended receiver while keeping it confidential from the other receiver. For this scenario, we offer an achievability scheme that combines the idea of using cooperative jamming with the Han-Kobayashi achievability scheme \cite{han}. More specifically, in our scheme each transmitter sends out a superposition of lattice codewords, taken from multiple nested lattice sets. The jamming codewords are also constructed using a lattice structure. Using careful alignments, each transmitter helps the other transmitter to keep its confidential message secret from the unintended receiver. This implies cooperation between the transmitters without any connection. To handle the finite SNR regimes, each receiver applies the compute-and-forward decoding strategy in \cite{ordentlich2014approximate}, \cite{ntranos2013usc}. We investigate the performance of our scheme for any finite SNR value (as long as $\log(SNR)>0$) and whenever the interference level lies either in the weak or moderately weak interference regimes. Also, we show that our achievable result reaches the optimal sum secure degrees of freedom for the considered model. Moreover, we provide a discussion on the extension of our scheme to the general case of $K>2$ users. \\
\indent The rest of the paper is organized as follows. In Section \ref{sec2}, we state the considered problem formally. In section \ref{sec3}, we present our achievable results. Section \ref{sec4} provides the proposed achievability scheme along with the analysis of security. Section \ref{sec5} extends our scheme to the $K>2$-user case. Finally, the paper is concluded in Section \ref{sec6}.
\section{Problem Statement}\label{sec2}
We consider the problem of reliable transmission over a two-user interference channel in which each transmitter has a confidential message to send to one intended receiver while keeping it secret from the other receiver. The relationships among channel inputs and outputs are described as:
\begin{IEEEeqnarray}{l}
\mathbf{y}_1=h_{11}\mathbf{x}_{1}+h_{21}\mathbf{x}_{2}+\mathbf{z}_1 \label{eqic1}\\
\mathbf{y}_2=h_{22}\mathbf{x}_{2}+h_{12}\mathbf{x}_{1}+\mathbf{z}_2 \label{eqic2}
\end{IEEEeqnarray} 
where $\mathbf{x}_{\ell}$ is transmitter $\ell$'s channel input with length $N$, and $\mathbf{y}_{\ell}$ is the channel output at receiver $\ell$, for $\ell=1,2$. The real-valued $h_{\ell \ell}$ is the channel gain from user $\ell$ to its respective receiver (the direct link gain) and the real-valued $h_{\ell j}$, for $j=1,2$ and $j\neq \ell$, is the cross link gain (the leakage link gain). We assume that the transmitters\footnote{In our scheme, knowledge of the channel state is not beneficial either to the receiver or to the eavesdropper.} know the channel states, i.e., the channel gains, in advance. Finally, the random vector $\mathbf{z}_{\ell}$ is an independent channel noise, which is i.i.d. Gaussian with zero mean and normalized variance.\\
\indent Transmitter~$\ell$ has an independent confidential message $W_{\ell}$, uniformly distributed over the set $\{1,\dots, 2^{NR_{\ell}}\}$, for $\ell \in~\{1,2\}$. Transmitter $\ell$ maps its message to codeword $\mathbf{x}_{\ell}$ through a stochastic encoder, i.e., $\mathbf{x}_{\ell} = \mathcal{E}_{\ell}(W_{\ell})$. Moreover, there is a power constraint on the channel input as $\|\mathbf{x_{\ell}}\|^2 \leq NP$, for $\ell=1,2$. At receiver $\ell$, decoder $\mathcal{D}_{\ell}$ estimates the respective transmitted message as $\hat{W}_{\ell}=\mathcal{D}_{\ell}(\mathbf{y}_{\ell})$. Figure 1 illustrates the communication model.
\begin{figure}\label{fig1}
\centering
\begin{tikzpicture}[scale=0.75]
\draw(0.7,6) node (nodew1) {$W_1$};
\draw (2,6) node[draw] (nodeE1) {$\mathcal E_1$};
\draw (7,6) node[circle,draw,inner sep=0] (nodeplus1) {$+$};
\draw (7,7) node (nodenoise1) {$\mathbf z_1$};
\draw (9,6) node[draw] (nodeD1) {$\mathcal D_1$};

\draw(10.7,6) node (nodewhat1) {$\begin{array}{c}
 \hat W_1 \end{array} $
 };
 \draw(11.5,6) node (nodewhat4) {$\begin{array}{c}
 W_2 \end{array} $
 }; 
\draw (11.5,6) node[cross=8pt,red] {};
\draw[->] (nodew1)--(nodeE1);
\draw[->] (nodeE1)--(nodeplus1) node[pos=0.1,sloped,above]{$\mathbf{x}_1$} node[pos=0.68,sloped,above] {$\scriptstyle \mathrm h_{11}$};
\draw[->] (nodenoise1)--(nodeplus1);
\draw[->] (nodeplus1)--(nodeD1) node[pos=0.3,sloped,above]{$\mathbf y_1$};
\draw[->] (nodeD1)--(nodewhat1);
\draw(0.7,1.5) node (nodew3) {$W_2$};
\draw (2,1.5) node[draw, minimum size = 10] (nodeE3) {$\mathcal E_2$};
\draw (7,1.5) node[circle,draw,inner sep=0] (nodeplus3) {$+$};
\draw (7,2.5) node (nodenoise3) {$ \mathbf{z}_2$};
\draw (9,1.5) node[draw] (nodeD3) {$ \mathcal D_2 $};
\draw(10.7,1.5) node (nodewhat2) {$\begin{array}{c}
 \hat W_2 \end{array} $
 };
\draw(11.5,1.5) node (nodewhat3) {$\begin{array}{c}
 W_1 \end{array} $
 }; 
\draw (11.5,1.5) node[cross=8pt,red] {};
\draw[->] (nodeD3)--(nodewhat2);
\draw[->] (nodew3)--(nodeE3);
\draw[->] (nodeE3)--(nodeplus3)node[pos=0.1,sloped,above]{$\mathbf{x}_2$} node[pos=0.36,sloped,above] {$\scriptstyle h_{22}$};
\draw[->] (nodenoise3)--(nodeplus3);
\draw[->] (nodeplus3)--(nodeD3) node[pos=0.3,sloped,above]{$\mathbf y_2$};
\draw[->] (3.25,6)--(nodeplus3) node[pos=0.2,below,yshift=-2]{$\scriptstyle \mathrm h_{12}$};
\draw[->] (3.25,1.5)--(nodeplus1) node[pos=0.6,above, yshift=3]{$\scriptstyle \mathrm h_{21}$};
\end{tikzpicture}
\vspace{0.02 in}
\caption{ \small {The two-user Gaussian interference channel model with confidential messages.}}
\vspace{-5mm}
\end{figure}
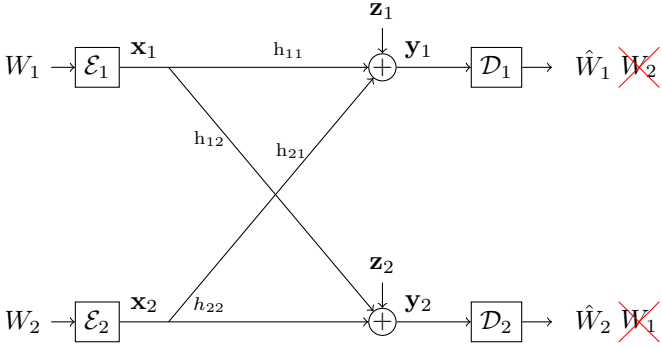
\begin{defi}[Achievable secure rates]\label{achievability tuple}
For the two-user Gaussian interference channel with independent confidential messages, a non-negative rate pair $(R_1,R_2)$ is achievable, if for any$~\epsilon>0$ and sufficiently large $N$, there exist encoders $\{\mathcal{E}_{\ell}\}_{\ell=1}^{2}$ and decoders $\{\mathcal{D}_{\ell}\}_{\ell=1}^{2}$ such that:
\begin{equation}
\mathrm{Prob}\left(D_{\ell}(\mathbf{y}_{\ell})\neq W_{\ell}\right)< \epsilon ~~ \ell=1,2\label{eqic3}
\end{equation}
\begin{equation}\label{eqic4}
R_{1}\leq \frac{1}{N}H(W_{1}|\mathbf{y}_2)+\epsilon,~~ 
R_{2}\leq \frac{1}{N}H(W_{2}|\mathbf{y}_1)+\epsilon
\end{equation}
\end{defi}
Inequalities in (\ref{eqic3}) capture the reliability constraints for both receivers and the constraints in (\ref{eqic4}) ensure the confidentiality of each message from the unintended receiver according to the notion of weak secrecy~\cite{maurer2000information}. The secrecy capacity region is the supremum over all the achievable secure rate pairs.
\begin{defi}[Weak and moderately weak interference regimes] In our model, the interference to noise ratio (INR) for receivers $1$ and $2$ is defined as\footnote{We assume that $\log(h_{11}^2P)>0$ and $\log(h_{22}^2P)>0$ which are consistent with the assumption that users operate above noise level.}
\begin{equation}
INR_{1}\triangleq h_{21}^2P,~~INR_{2}\triangleq h_{12}^2P
\end{equation}
Then, the \textit{weak interference regime} includes all channel gains such that
\begin{equation}\label{eqic6}
\frac{1}{2}\leq\left(\frac{\log(h_{21}^2P)}{\log(h_{11}^2P)}\right)<\frac{2}{3},~~\frac{1}{2}\leq\left(\frac{\log(h_{12}^2P)}{\log(h_{22}^2P)}\right)<\frac{2}{3}
\end{equation}
Furthermore, the \textit{moderately weak interference regime} includes all channel gains such that 
\begin{equation}\label{eqic7}
\frac{2}{3}\leq\left(\frac{\log(h_{21}^2P)}{\log(h_{11}^2P)}\right)<1,~~\frac{2}{3}\leq\left(\frac{\log(h_{12}^2P)}{\log(h_{22}^2P)}\right)<1
\end{equation}
\end{defi}
\indent Note that Definition 2 is aligned with the common notions of weak and moderately weak interference regimes in the literature, e.g., as in \cite{ordentlich2014approximate}.
\section{Main Results}\label{sec3}
Consider $P_{t1},P_{u1},P_{t2},P_{u2}$ as non-negative scalar variables such that $P_{t1}+\left(\frac{h_{21}}{h_{11}}\right)^2P_{u1}\leq P$ and $P_{t2}+\left(\frac{h_{12}}{h_{22}}\right)^2P_{u2}\leq P$. Then, the following theorem provides a lower bound on the achievable secure rates.
\begin{theorem}
A rate pair $(R_1,R_2)$ which satisfies the following inequalities is an achievable secure rate pair for the weak and moderately weak interference regimes.
\begin{IEEEeqnarray}{l}
R_1<R_{comb,2}^{(1)}-\frac{1}{2}\log\left(\frac{P_{u2}+P_{t1}}{P_{u2}}\right)\label{eqic10}\\
R_2<R_{comb,2}^{(2)}-\frac{1}{2}\log\left(\frac{P_{u1}+P_{t2}}{P_{u1}}\right)\label{eqic11}
\end{IEEEeqnarray}
In which, $R_{comb,2}^{(1)}$ is a lower bound on the optimal combination rate at which transmitter 1's message can be reliably decoded by receiver 1 using the asymmetric compute-and-forward decoding strategy. The achievable combination rate $R_{comb,2}^{(1)}$ is mathematically computed in (\ref{eqic24}). The combination rate $R_{comb,2}^{(2)}$ is similarly defined for receiver 2. The proof of Theorem 1 is shown in Section~\ref{sec4}. 
\end{theorem}
\begin{corollary} The secure rates in (\ref{eqic10}) and (\ref{eqic11}) scale linearly with $\log(P)$.
\end{corollary}
The proof of Corollary 1 is provided in Section~\ref{sec4}.
\begin{corollary} The optimal sum secure degrees of freedom of $\frac{2}{3}$ is achievable using our proposed scheme, i.e.,
\begin{equation}
\lim_{P\rightarrow \infty} \frac{R_1+R_2}{\frac{1}{2}\log(1+P)}= \frac{2}{3}
\end{equation}
\end{corollary}
The achievability proof of Corollary 2 is deduced by applying Corollary 5 in \cite{ordentlich2014approximate} to $R_{comb,2}^{(1)}$ and $R_{comb,2}^{(2)}$ and the fact that the second terms in (\ref{eqic10}) and (\ref{eqic11}) are constant with respect to power $P$. Also, the upper bound was shown in \cite{xie2014secure}.
\begin{remark} Recall that the performance of i.i.d. Gaussian random codes was investigated in \cite{liu2008discrete} under three different schemes including time-sharing, multiplexed transmission, and incorporation of artificial noise. According to the results in \cite{liu2008discrete}, i.i.d. Gaussian random codes achieve \textit{zero} sum secure degrees of freedom for the two-user Gaussian interference channel in all the three schemes.
\end{remark}
\section{Achievability Scheme}\label{sec4}
We begin the achievabiliy proof of Theorem 1 by describing our codebook construction, encoding, decoding strategies, and analysis of security.
\subsection{Codebook construction}
Our codebook construction is motivated by the Han-Kobayashi scheme in which each user transmits a superposition of codewords taken from two nested lattice codebooks. However, the difference here is that the second lattice codebook is used to encode the jamming signal. Therefore, each transmitter encodes its message as well as a cooperative jamming signal that masks the other transmitter's message at the unintended receiver. We describe the codebook construction for transmitter 1; transmitter 2 builds its codebook similarly. The transmitter picks two $n$-dimensional pairs of coarse and fine lattice sets $(\Lambda_{t,1},\Lambda_{t,f,1})$ and $(\Lambda_{u,1},\Lambda_{u,f,1})$. The former coarse and fine lattice pair, indexed by $t$, is used for encoding transmitter's message and the latter pair indexed by $u$ is used for encoding the transmitter's jamming signal. Assume that the lattice sets chosen by both transmitters form a nested structure as\footnote{$\Lambda$ is the common coarsest lattice set.}
\begin{equation}\label{eqic12}
\medmuskip=0mu
\thinmuskip=0mu
\thickmuskip=0mu
\Lambda\subseteq \Lambda_{t,2}\subseteq \Lambda_{t,1}\subseteq \Lambda_{u,2}\subseteq \Lambda_{u,1}\subseteq\Lambda_{t,f,2}\subseteq \Lambda_{t,f,1}\subseteq \Lambda_{u,f,2}\subseteq \Lambda_{u,f,1}
\end{equation}
We scale the coarse lattice sets such that the second moments of $\Lambda_{t,2}, \Lambda_{t,1}, \Lambda_{u,2}, \Lambda_{u,1}$ are determined by $P_{t,2},P_{t,1},P_{u,2},P_{u,1}$, respectively. We denote the fundamental Voronoi region of the coarse lattice $\Lambda_{t,1}$ as $\mathcal{V}_{t,1}$. The centers of the cosets of the fine lattice $\Lambda_{t,f,1}$ are $n$-length lattice words which are considered as the realizations of the $n$-length random vector $\mathbf{t}_{1}$. Then, the lattice codebook of transmitter 1 is constructed as $\mathcal{L}_{t,1}\triangleq\{\mathbf{t}_{1}|\mathbf{t}_{1}\in \mathcal{V}_{t,1}\}$ and is dubbed as the \textit{inner} codebook. Assume a probability distribution $P(\mathbf{t}_{1})$ over the inner codebook $\mathcal{L}_{t,1}$. Then, transmitter 1 constructs a realization of an $N$-length random vector $\bar{\mathbf{t}}_{1}$, where $N\triangleq n \times B$, by generating $B$ i.i.d. copies of $ \mathbf{t}_{1}$ according to the distribution $P(\mathbf{t}_{1})$. This process is repeated for $2^{NR_{comb,2}^{(1)}}$ times. The collection of the generated vectors are called as the \textit{outer} codebook and is denoted by $\mathcal{C}_{t,1}$. Also, vector $\bar{\mathbf{t}}_{1}$ represents the random vector for the \textit{outer} lattice codewords of transmitter 1. Note that $R_{comb,2}^{(1)}\triangleq \frac{1}{n}\log(|\mathcal{L}_{t,1}|)$ is the rate at which the inner codewords of transmitter 1 are generated.\\
\indent In addition to the codebooks for the confidential message, the transmitter constructs an inner codebook $\mathcal{L}_{u,1}$ and an outer codebook  $\mathcal{C}_{u,1}$ for the \textit{jamming} signal, in a similar manner. The random vector assigned to the jamming inner lattice codewords is denoted by $\mathbf{u}_{1}$, and $\bar{\mathbf{u}}_1$ represents the random vector for the jamming outer codewords. In the next step, codebook $\mathcal{C}_{t,1}$ is randomly partitioned into $2^{NR_1}$ bins of equal size. The random variable representing the bin index is denoted by $W_1$ and its realization $w_1$ takes values from the set $\{1,\dots,2^{NR_1}\}$. $\mathcal{C}_{t,1}(w_1)$ is the set of outer codewords belong to bin $w_1$. \\
\indent Finally, a set of random $N$-length vectors $\bar{\mathbf{d}}_{t,1}$ and $\bar{\mathbf{d}}_{u,1}$ are generated for dithering. Assume that each $n$-length block of dither $\bar{\mathbf{d}}_{t,1}$ has a uniform distribution over $\mathcal{V}_{t,1}$, and similarly, each $n$-length block of dither $\bar{\mathbf{d}}_{u,1}$ has a uniform distribution over the corresponding Voronoi region $\mathcal{V}_{u,1}$. Dithers are public and hence they don't add to secrecy. Note that transmitter 2 constructs its codebooks similarly.
\subsection{Encoding}
We describe the encoding procedure for transmitter 1; similar arguments hold for encoding at transmitter 2. To encode the confidential message $w_1$, transmitter 1 randomly picks a codeword $\bar{\mathbf{t}}_1$ from the bin set $\mathcal{C}_{t,1}(w_1)$. Then, it dithers the codeword using a randomly generated $N$-length vector $\bar{\mathbf{d}}_{t,1}$. The result is reduced to the Voronoi region $\mathcal{V}_{t,1}$ using the modular operation. Also, $\bar{\mathbf{u}}_1$ is chosen at random. We have:
\begin{equation}\label{eqic13}
\medmuskip=0mu
\thinmuskip=0mu
\thickmuskip=0mu \bar{\mathbf{t}}_{1,d}\triangleq\left[\bar{\mathbf{t}}_1+\bar{\mathbf{d}}_{t,1}\right]\modvf,~\bar{\mathbf{u}}_{1,d}\triangleq\left[\bar{\mathbf{u}}_1+\bar{\mathbf{d}}_{u,1}\right]\moduf
\end{equation}
Next step in the encoding procedure is scaling the jamming codeword such that it aligns with the confidential codeword of transmitter 2 at receiver 1. The superposition of the confidential codeword with the scaled jamming codeword is sent through the channel as the transmitter 1 input, i.e., $\mathbf{x}_1=\bar{\mathbf{t}}_{1,d}+\frac{h_{21}}{h_{11}}\bar{\mathbf{u}}_{1,d}$. Note that $\mathbf{x}_1$ satisfies the channel input power constraint, thus, $P_{t1}+\left(\frac{h_{21}}{h_{11}}\right)^2P_{u1} \leq P$.
\subsection{Decoding}
As in the previous steps, we describe the decoding procedure at receiver 1; receiver 2 acts similarly. The decoding procedure is based on the asymmetric compute-and-forward strategy, introduced in \cite{ordentlich2014approximate} and \cite{ntranos2013usc}. In our model, receiver 1 observes sequence $\mathbf{y}_1$ from the channel as
\begin{equation}\label{eqic15}
\mathbf{y}_1=h_{11}\bar{\mathbf{t}}_{1,d}+h_{21}(\bar{\mathbf{t}}_{2,d}+\bar{\mathbf{u}}_{1,d})+\frac{h_{21}h_{12}}{h_{22}}\bar{\mathbf{u}}_{2,d}+\mathbf{z}_1
\end{equation}
Assume that the powers of codewords $\bar{\mathbf{u}}_{1,d}$ and $\bar{\mathbf{u}}_{2,d}$ are set such that the codewords $\frac{h_{12}h_{21}}{h_{11}}\bar{\mathbf{u}}_{1,d}$ and $\frac{h_{21}h_{12}}{h_{22}}\bar{\mathbf{u}}_{2,d}$ are below the noise power level. As jamming codewords do not carry useful information about the confidential signals, receiver 1 treats the third term in (\ref{eqic15}) as noise. Therefore, receiver 1 observes an effective two-user Gaussian multiple-access channel (GMAC) as $\mathbf{y}_1=h_{11}\bar{\mathbf{t}}_{1,d}+h_{21}(\bar{\mathbf{t}}_{2,d}+\bar{\mathbf{u}}_{1,d})+\tilde{\mathbf{z}}_1$, in which $\tilde{\mathbf{z}}_1$ is the effective noise seen by receiver 1. Now, let us normalize the noise power to form a standard effective MAC as in \cite{ordentlich2014approximate}. We have:
\begin{equation}\label{eqic17}
\tilde{\mathbf{y}}_1=\frac{h_{11}}{\sqrt{1+\alpha_1^2P_{u2}}}\bar{\mathbf{t}}_{1,d}+\frac{h_{21}}{\sqrt{1+\alpha_1^2P_{u2}}}(\bar{\mathbf{t}}_{2,d}+\bar{\mathbf{u}}_{1,d})+\tilde{\mathbf{z}}_{eff,1}
\end{equation} 
where $\tilde{\mathbf{z}}_{eff,1}$ is the normalized-power effective noise; the factor $\alpha_1$ is defined as $\alpha_1\triangleq \frac{h_{12}h_{21}}{h_{22}}$; finally, $\tilde{\mathbf{y}}_1$ is the scaled sequence observed at receiver 1. \\
\indent Consider the effective channel vector, $\mathbf{h}_{eff,1}$, and the power scaling vector, $\mathbf{b}_{eff,1}$, defined as $\mathbf{h}_{eff,1}\triangleq\left(\frac{h_{11}}{\sqrt{1+\alpha_1^2P_{u2}}},\frac{h_{21}}{\sqrt{1+\alpha_1^2P_{u2}}}\right)^T$ and $\mathbf{b}_{eff,1}\triangleq \left(\sqrt{\frac{P_{t1}}{P}},\sqrt{\frac{P_{t2}+P_{u1}}{P}}\right)^T$, respectively. Then, according to Theorem 7 in~\cite{ordentlich2014approximate}, the sum of the optimal combination rates for the above effective two-user GMAC seen by receiver 1 is lower-bounded as
\begin{equation}\label{eqic20}
\medmuskip=0mu
\thinmuskip=0mu
\thickmuskip=0mu
\sum_{\ell=1}^2R_{comb,\ell}^{(1)}\geq \frac{1}{2}\log\left(\frac{1+P\sum_{\ell=1}^2 \mathbf{h}_{eff,1}^2(\ell) \mathbf{b}_{eff,1}^2(\ell)}{\mathbf{b}_{eff,1}^2(1)\mathbf{b}_{eff,1}^2(2)}\right)-1
\end{equation}
Next, assume the optimal combination rates for the aforementioned effective GMAC are sorted in a descending order, i.e., $R_{comb,1}^{(1)}\geq R_{comb,2}^{(1)}$. Then, based on Theorem 9 in~\cite{ordentlich2014approximate}, it is guaranteed that transmitter 1's confidential message can be decoded reliably by receiver 1 for all rates not bigger than $R_{comb,2}^{(1)}$ unless the effective channel gains are rational\footnote{The Lebesgue measure of such event is small \cite{ordentlich2014approximate}}. As a result, it remains to find a lower bound on the achievable rate $R_{comb,2}^{(1)}$, i.e.,
\vspace{-0.4 cm}
\begin{equation}\label{eqic21}
R_{comb,2}^{(1)}=\sum_{\ell=1}^2R_{comb,\ell}^{(1)}-R_{comb,1}^{(1)}
\end{equation}
which is equivalent to finding an upper bound on the achievable rate $R_{comb,1}^{(1)}$, which maps to the decoding rate of the aligned codewords $\bar{\mathbf{t}}_{2,d}+\bar{\mathbf{u}}_{1,d}$. Therefore, the computation rate for $R_{comb,1}^{(1)}$ is given as $R_{comb,1}^{(1)}=\frac{1}{2}\log(P_{t2}+P_{u1})-\frac{1}{2}\log(\sigma_{eff,1}^2)$, in which $\sigma_{eff,1}^2$ is the variance of the effective noise in \textit{the first integer linear combination of the codewords decoded at receiver 1}. Let us denote the first integer linear combination as $\mathbf{v}_1$. Assume it is determined by the integer-valued $2\times 1$ vector $\mathbf{a}_1$, i.e., $\mathbf{v}_1\triangleq\mathbf{a}_1(1)(\bar{\mathbf{t}}_2+\bar{\mathbf{u}}_1)+\mathbf{a}_1(2)\bar{\mathbf{t}}_1$. Recall that according to the compute-and-forward strategy in \cite{nazer2011compute}, the receiver decodes an estimate of  $\mathbf{v}_1$ as follows:
\begin{equation}\label{eqic22nim1}
\mathbf{s}_1=\left[\beta\tilde{\mathbf{y}}_1-\mathbf{a}_1(1)(\bar{\mathbf{d}}_{t,2}+\bar{\mathbf{d}}_{u,1})-\mathbf{a}_1(2)\bar{\mathbf{d}}_{t,1}\right]\modn
\end{equation}
\begin{equation}\label{eqic22nim2}
=\left[\mathbf{v}_1+\mathbf{z}_{eff,1}\right]\modn
\end{equation}
in which $\beta \in \mathbb{R}$ is a scaling factor. To estimate $\mathbf{v}_1$ from (\ref{eqic22nim2}), the receiver quantizes (\ref{eqic22nim2}) with respect to the finest participating lattice, i.e., $\hat{\mathbf{v}}\triangleq Q_{\Lambda_{u,f,1}}(\mathbf{s}_1)$. Note that the modular operation as well as the quantization are done block-wise.\\
\indent To upper-bound the combination rate $R_{comb,1}^{(1)}$, it is sufficient to lower bound $\sigma_{eff,1}^2$; which is computed as
\begin{equation*}
\medmuskip=-1mu
\thinmuskip=-1mu
\thickmuskip=-1mu
\sigma_{eff,1}^2\triangleq\left(\beta^2+P\left(\mathbf{b}_{eff,1}^2(1)\left(\beta\mathbf{h}_{eff,1}(1)-\mathbf{a}_1(1)\right)^2\right.\right.
\end{equation*}
\begin{equation}\label{eqic23}
\medmuskip=-1mu
\thinmuskip=-1mu
\thickmuskip=0mu
\left. \left. + \mathbf{b}_{eff,1}^2(2)\left(\beta\mathbf{h}_{eff,1}(2)-\mathbf{a}_1(2)\right)^2\right)\right)
\end{equation}
\indent Finally, the effective variance is minimized over $\beta$ and $\mathbf{a}_1$ and it is denoted by $\sigma_{eff,1}^{*2}$, i.e., $\sigma_{eff,1}^{*2}\triangleq \min_{\substack{\beta,\mathbf{a}}}\sigma_{eff,1}^2$. As a result, a lower bound on the achievable rate $R_{comb,2}^{(1)}$ is obtained as
\begin{IEEEeqnarray}{l}
\nonumber R_{comb,2}^{(1)}\geq \frac{1}{2}\log\left(1+\mathbf{h}_{eff,1}^2(1)P_{t1}+\mathbf{h}_{eff,2}^2(2)\left(P_{t2}+P_{u1}\right)\right) \\ \label{eqic24}
\medmuskip=-1mu
\thinmuskip=-1mu
\thickmuskip=-1mu
+\log(P)+ \frac{1}{2}\log (\sigma_{eff,1}^{*2})
-\frac{1}{2}\log (P_{t1})-\log(P_{t2}+P_{u1})-1
\end{IEEEeqnarray}
\indent Prior to security analysis, we show that the rates in (\ref{eqic10}) and (\ref{eqic11}) scale with power as it was claimed in Corollary 1.\\
\indent \textit{Proof of Corollary 1:} We show that $R_1$ scales linearly with $\log(P)$; the same result can be shown for $R_2$, using similar arguments. To this end, let us assume a power allocation among the users' jamming powers and confidential-messages powers as $P_{t1}=P_{t2}=(1-\gamma^2)P$ and $(\frac{h_{21}}{h_{11}})^2P_{u1}=(\frac{h_{12}}{h_{22}})^2P_{u2}=\gamma^2P$, for some $0<\gamma^2<1$. This is a valid choice as it satisfies the power constraint. Substituting these power values in (\ref{eqic10}), we observe that the second term in (\ref{eqic10}) is constant with respect to P for any valid choice of $\gamma$. Therefore, it is enough to show that $R_{comb,2}^{(1)}$ grows linearly with $\log(P)$. Now assume $\gamma^2=\frac{1}{h_{21}^2P}$, which may be a sub-optimal choice. The validity of this choice can be checked easily for the weak and the moderately weak interference regimes. Note that this choice for $\gamma^2$ makes the power of the third term in (\ref{eqic15}) to be within noise level. Having chosen $\gamma^2$ as mentioned, we can compute $P_{u2},P_{t1}$ accordingly. As a result, we observe that the first and fourth terms in (\ref{eqic24}) can be rewritten as $\frac{1}{2}\log(P)+c_1$ and $\frac{1}{2}\log(P)+c_2$ for some constants (with respect to P) $c_1,c_2$, respectively; additionally, the fifth term can be simplified as $\log(P)+c_3$ for a constant $c_3$. As a result, we have $R_{comb,2}^{(1)}\geq \frac{1}{2}\log(\sigma_{eff,1}^{*2})+c_1+c_2+c_3$. Now, consider the expression in (\ref{eqic23}); note that it can be rewritten as $\sigma_{eff,1}^2=(\beta^2+c_4(\mathbf{a},\beta)P)$ in which $c_4(.)$ is a positive number. Since the latter holds for any choice of $\beta$ and integer coefficient vector $\mathbf{a}$, it is also true for the the infimum and as a result, $\log(\sigma_{eff,1}^{*2})\propto \log(P)$. This completes the proof of Corollary 1.
\subsection{Analysis of Security}
In this subsection we show that our scheme provides weak secrecy for each transmitter's confidential message. Specifically, we prove weak secrecy for transmitter 1's confidential message $W_1$; the proof of weak secrecy for transmitter 2's message is deduced similarly. We have
\begin{IEEEeqnarray}{l}
\medmuskip=-1mu
\thinmuskip=-1mu
\thickmuskip=0mu
\frac{1}{N}I(W_1;\mathbf{y}_2)\leq \frac{1}{N}I(W_1;\mathbf{y}_2,\bar{\mathbf{t}}_2)=R_1- \frac{1}{N}H(W_1|\mathbf{y}_2,\bar{\mathbf{t}}_2) \label{eqic25}
\end{IEEEeqnarray}
Next, we find a lower bound on the second term in (\ref{eqic25}) as follows:
\begin{IEEEeqnarray}{l}
\nonumber  \frac{1}{N}H(W_1|\mathbf{y}_2,\bar{\mathbf{t}}_2)
= \frac{1}{N}H(\bar{\mathbf{t}}_1,W_1|\mathbf{y}_2,\bar{\mathbf{t}}_2)-\frac{1}{N}H(\bar{\mathbf{t}}_1|\mathbf{y}_2,\bar{\mathbf{t}}_2,W_1)\\\nonumber
\geq \frac{1}{N}H(\bar{\mathbf{t}}_1|\mathbf{y}_2,\bar{\mathbf{t}}_2)-\frac{1}{N}H(\bar{\mathbf{t}}_1|\mathbf{y}_2,\bar{\mathbf{t}}_2,W_1)\\
\medmuskip=-1mu
\thinmuskip=-1mu
\thickmuskip=-1mu
\stackrel{(a)}{\geq}\frac{1}{N}H(\bar{\mathbf{t}}_1|\mathbf{y}_2,\bar{\mathbf{t}}_2)-2\epsilon_2
~\stackrel{(b)}{\geq}\frac{1}{N}H(\bar{\mathbf{t}}_1|\mathbf{y}_2,\bar{\mathbf{t}}_2,\bar{\mathbf{u}}_1,\mathbf{z}_2,D)-2\epsilon_2 \label{eqic26}
\end{IEEEeqnarray}
Inequality (a) is deduced by applying Lemma 1 in \cite{babaheidarian2} to outer codewords $\bar{\mathbf{t}}_1$. Inequality (b) holds since conditioning reduces entropy. In (\ref{eqic26}), $D$ denotes the collection of all the dither vectors. Note that receiver 2 observes $\mathbf{y}_2=h_{22}\bar{\mathbf{t}}_{2,d}+h_{12}(\bar{\mathbf{t}}_{1,d}+\bar{\mathbf{u}}_{2,d})+\frac{h_{12}h_{21}}{h_{11}}\bar{\mathbf{u}}_{1,d}+\mathbf{z}_2$. Hence, if the receiver 2 had the information of the random vectors $D$, $\mathbf{z}_2$, and $\bar{\mathbf{t}}_2$, it could decode the aligned lattice codeword $\bar{\mathbf{t}}_{1,d}+\bar{\mathbf{u}}_{2,d}$. As a result, based on (\ref{eqic26}), we have
\begin{IEEEeqnarray}{l}
\medmuskip=-1mu
\thinmuskip=-1mu
\thickmuskip=-1mu
\nonumber  \frac{1}{N}H(W_1|\mathbf{y}_2,\bar{\mathbf{t}}_2)
\geq\frac{1}{N}H(\bar{\mathbf{t}}_1|\bar{\mathbf{t}}_{1,d}+\bar{\mathbf{u}}_{2,d},\bar{\mathbf{t}}_2,\bar{\mathbf{u}}_1,\mathbf{z}_2,D)-2\epsilon_2\\
\nonumber
\medmuskip=-1mu
\thinmuskip=-1mu
\thickmuskip=-1mu
=\frac{1}{N}H\left(\bar{\mathbf{t}}_1\bigg|\left[\bar{\mathbf{t}}_{1,d}+\bar{\mathbf{u}}_{2,d}\right] \modus, Q_{\Lambda_{u,2}}(\bar{\mathbf{t}}_{1,d}+\bar{\mathbf{u}}_{2,d}),\bar{\mathbf{t}}_2,\bar{\mathbf{u}}_1,\mathbf{z}_2,D\right)\\\nonumber -2\epsilon_2\\\nonumber
\medmuskip=-1mu
\thinmuskip=-1mu
\thickmuskip=-1mu
\geq \frac{1}{N}H\left(\bar{\mathbf{t}}_1\bigg|\left[\bar{\mathbf{t}}_{1,d}+\bar{\mathbf{u}}_{2,d}\right] \modus,\bar{\mathbf{t}}_2,\bar{\mathbf{u}}_1,\mathbf{z}_2,D\right)\\
\nonumber
\medmuskip=-1mu
\thinmuskip=-1mu
\thickmuskip=-1mu
-\frac{1}{N}H(Q_{\Lambda_{u,2}}(\bar{\mathbf{t}}_{1,d}+\bar{\mathbf{u}}_{2,d})\big|\bar{\mathbf{t}}_2,\bar{\mathbf{u}}_1,\mathbf{z}_2,D)-2\epsilon_2\\
\nonumber
\medmuskip=-1.5mu
\thinmuskip=-1.5mu
\thickmuskip=-1.5mu
\stackrel{(a)}{=} \frac{1}{N}H\left(\bar{\mathbf{t}}_1\bigg|\left[\bar{\mathbf{t}}_1+\bar{\mathbf{u}}_2\right] \modus\right)
\end{IEEEeqnarray}
\begin{IEEEeqnarray}{l}
\nonumber
\medmuskip=-1.5mu
\thinmuskip=-1.5mu
\thickmuskip=-1.5mu
-\frac{1}{N}H\left(Q_{\Lambda_{u,2}}(\bar{\mathbf{t}}_{1,d}+\bar{\mathbf{u}}_{2,d})\bigg|\bar{\mathbf{t}}_2,\bar{\mathbf{u}}_1,\mathbf{z}_2,D\right)-2\epsilon_2\\
\nonumber
\medmuskip=-1mu
\thinmuskip=-1mu
\thickmuskip=-1mu
\stackrel{(b)}{=}\frac{1}{N}H(\bar{\mathbf{t}}_1)-\frac{1}{N}H(Q_{\Lambda_{u,2}}(\bar{\mathbf{t}}_{1,d}+\bar{\mathbf{u}}_{2,d})\big|\bar{\mathbf{t}}_2,\bar{\mathbf{u}}_1,\mathbf{z}_2,D)-2\epsilon_2\\\nonumber
\medmuskip=-1mu
\thinmuskip=-1mu
\thickmuskip=-1mu
\geq \frac{1}{N}H(\bar{\mathbf{t}}_1)-\frac{1}{N}H(Q_{\Lambda_{u,2}}(\bar{\mathbf{t}}_{1,d}+\bar{\mathbf{u}}_{2,d}))-2\epsilon_2\\
\nonumber
\medmuskip=-1mu
\thinmuskip=-1mu
\thickmuskip=-1mu
\stackrel{(c)}{\geq}\frac{1}{N}H(\bar{\mathbf{t}}_1)
-\frac{1}{2}\log\left(\frac{P_{u2}+P_{t1}}{P_{u2}}\right)-\delta(\epsilon)-2\epsilon_2\\
\medmuskip=-1mu
\thinmuskip=-1mu
\thickmuskip=-1mu
=R_{comb,2}^{(1)}-\frac{1}{2}\log\left(\frac{P_{u2}+P_{t1}}{P_{u2}}\right)-\delta(\epsilon)-2\epsilon_2 \label{eqic28}
\end{IEEEeqnarray}
in which equality (a) is due to the fact that the random vectors $\bar{\mathbf{t}}_1, \bar{\mathbf{u}}_2$ are independent from the dithers, the noise and random vectors $\bar{\mathbf{u}}_1, \bar{\mathbf{t}}_2$. Equality (b) follows from Crypto Lemma, Lemma 2 in \cite{forney2003role}, which states that the lattice codeword $\left[\bar{\mathbf{t}}_1+\bar{\mathbf{u}}_2\right] \modus$ belongs to the codebook $\mathcal{L}_{u,2}$ (operation mod is done for each $n$-length block) and is independent of codeword $\bar{\mathbf{t}}_1$. Finally, inequality (c) is deduced from Lemma 1 in~\cite{paper1}, which bounds the discrete entropy of the quantized vector. Eventually, from (\ref{eqic10}), (\ref{eqic28}), and (\ref{eqic25}), the weak secrecy proof for transmitter 1's confidential message is concluded.
\section{The general case: The Gaussian Interference Channel with $K>2$ users} \label{sec5}
Our scheme in Section~\ref{sec4} can be modified to preserve the confidentiality of all the messages from the unintended receivers when there are $K>2$ users. Recall that in our scheme, each jamming signal was designed to protect one confidential message at one receiver. For the case of $K>2$ users, we divide each confidential message into $K-1$ independent random sub-messages and assign each of them a lattice codebook. For transmitter $\ell$, the sub-messages outer codewords are denoted by $\{\bar{\mathbf{t}}_{\ell,i}\}_{i=1,i\neq \ell}^K$; for $\ell\in \{1,\dots, K\}$. Now, each jamming codeword protects a portion of all sub-messages at all the required receivers simultaneously. For instance, the jamming codeword of transmitter $\ell$, i.e., $\bar{\mathbf{u}}_\ell$ conceals all codewords $\{\bar{\mathbf{t}}_{i,\ell}\}_{i=1,i\neq \ell, i \neq j}^K$ at receiver $j$. However, this requires that $\bar{\mathbf{u}}_\ell$ get aligned with the same codeword at multiple receiver even though the channel link gains are not the same for different receivers. As an example, consider the case of $K=3$: $\bar{\mathbf{u}}_1$ needs to align with $\bar{\mathbf{t}}_{2,1}$ at receivers 1 and 2; additionally, it should protect $\bar{\mathbf{t}}_{3,1}$ at receivers 1 and 3. Clearly, perfect alignment would not be achieved given that channel link gains are not the same. To remedy this alignment issue, we incorporate a generalization of the asymptotic \textit{real} alignment proposed in \cite{MotahariRef16} and used in \cite{xieJournal2}, \cite{khisti2011interference}. Using this technique, we further split each sub-message inner codeword into large number of components each of which is an $n$-dimensional lattice vector. Next, using a proper beamforming of the transmitted signals, we can show that a subset of components of each codeword gets aligned with a subset of components of a jamming codeword. Hence, even though a perfect alignment between two desired codewords cannot be achieved at more than one receiver, a partial alignment among their corresponding components can occur at all the required receivers, simultaneously. It can be shown that for a large number of components, the desired alignments happen at all the receivers asymptotically. We aim to further elaborate our scheme for this general case in the extended version of this paper.
\section{Conclusion}\label{sec6}
In this paper, we considered a reliable and secure communication scenario through the two-user Gaussian interference channel when the interference level is either weak or moderately weak. We showed that our achievable result scales linearly with $\log(SNR)$, when $\log(SNR)>0$, and reaches the optimal sum secure degrees of freedom at infinite SNR. We also argued how our scheme could be extended to the $K>2$-user case. 
\bibliography{refn}
\end{document}